\documentclass[10pt,conference]{IEEEtran}
\IEEEoverridecommandlockouts
\usepackage{cite}
\usepackage{amsmath,amssymb,amsfonts}
\usepackage{algorithmic}
\usepackage{graphicx}
\usepackage{textcomp}
\usepackage{svg}
\usepackage[obeyspaces]{url}
\usepackage{xcolor}
\usepackage{url}
\usepackage{graphicx}
\usepackage{ragged2e}
\usepackage{hyperref}
\usepackage{booktabs}
\usepackage{subcaption}
\usepackage{diagbox}
\usepackage[ruled,linesnumbered]{algorithm2e}
\SetKw{KwBy}{by}
\usepackage{mathtools}
\DeclarePairedDelimiter\ceil{\lceil}{\rceil}

\def\BibTeX{{\rm B\kern-.05em{\sc i\kern-.025em b}\kern-.08em
    T\kern-.1667em\lower.7ex\hbox{E}\kern-.125emX}}

\usepackage[utf8]{inputenc}
\usepackage[english]{babel}
\usepackage{amsthm}
\usepackage{mathrsfs}
\newcommand{\etal}{\textit{et al}.~}

\usepackage{lastpage}
\usepackage{fancyhdr}

\usepackage{amsmath}

\theoremstyle{definition}

\begin{document}

\title{Real-Time Systems Optimization with Black-box Constraints and Hybrid Variables}


\author{\IEEEauthorblockN{
Sen Wang$^1$,
Dong Li$^1$,
Shao-Yu Huang$^2$,
Xuanliang Deng$^{1}$,
Ashrarul H. Sifat$^{1}$, \\
Changhee Jung$^2$, Ryan Williams$^1$, Haibo Zeng$^1$
\thanks{For any inquiries or questions, please contact $\{$swang666, hbzeng$\}$@vt.edu
}
}
\IEEEauthorblockA{\textit{$^1$Virginia Tech, $^2$Purdue University}} 
\vspace{-10mm}
}
\maketitle

\thispagestyle{plain}
\pagestyle{plain}

\begin{abstract}

When optimizing real-time systems, designers often face a challenging problem where the schedulability constraints are non-convex, non-continuous, or lack an analytical form to understand their properties. 
Although the optimization framework NORTH proposed in \cite{Wang2023RTAS} is general (it works with arbitrary schedulability analysis) and scalable, it can only handle problems with continuous variables, which limits its application.
In this paper, we extend the applications of the framework NORTH to problems with a hybrid of continuous and discrete variables. This is achieved in a coordinate-descent method, where the continuous and discrete variables are optimized separately during iterations. The new framework, NORTH+, improves around 20\% solution quality than NORTH in experiments.
\end{abstract}

\section{Introduction}
Over the years, the real-time systems community has developed an impressive list of scheduling algorithms and schedulability analysis techniques. However, many schedulability analyses are not friendly for optimization purposes. Some analysis methods are complicated and do not provide closed-form expressions, such as those based on demand bound functions~\cite{Baruah1990PreemptivelySH}, real-time calculus~\cite{Thiele2000}, abstract event model interfaces~\cite{henia2005system}, or timed automata~\cite{Larsen1997UppaalIA, Nasri2019ResponseTimeAO}.

Another challenge in optimizing real-time systems is their ever-increasing complexity.  
The functionality of real-time systems such as automotive and aircraft is growing fast, especially given the recent trend of introducing autonomous features and system-to-system connectivity \cite{Heintzman2021-mw, Percept21_RTSS}. 
The existing optimization algorithms do not scale to the hundreds of software tasks that modern automotive systems contain~\cite{Kramer15benchmark}.

The existing approaches for optimizing real-time systems do not adequately address the above two challenges as they lack applicability or scalability. Besides, many standard mathematical optimization frameworks cannot efficiently handle complicated schedulability analysis constraints. 
Therefore, in recent work~\cite{Wang2023RTAS}, Wang \etal introduced \textit{NORTH} (Numerical Optimizer with Real-Time Highlight), an optimization framework for optimizing systems with black-box schedulability constraints based on numerical optimization algorithms. 
However, the application of NORTH is limited to systems with only continuous variables.

In this paper, we extend NORTH~\cite{Wang2023RTAS} to handle problems with \textit{hybrid} (i.e., a mixture of continuous and discrete) variables.
The new framework, named NORTH+, achieves this extension based on an iterative algorithm. During the iterations, the continuous and discrete variables are optimized independently. The continuous optimization component remains the same as NORTH, while the discrete optimization can leverage heuristics to ensure algorithm scalability. 
Simulated experiments indicate that NORTH+ brings around 20$\%$ performance improvements compared with NORTH~\cite{Wang2023RTAS}.

\section{Problem description}
Generally speaking, real-time systems design can be mathematically described as an optimization problem:
\begin{align}
    \min\limits_{\textbf{x}} \ \ \ \ \ \  & \mathcal{F}(\textbf{x}) 
    \label{general_F} \\
    \text{subject to}  \ \ \  &   \text{Sched}(\textbf{x}) = 0 \label{schedulability_analysis_true_false} 
\end{align}
where the vector of design variables $\textbf{x}=\{\textbf{x}^{\mathcal{C}}, \textbf{x}^{\mathcal{D}} \}$ includes continuous variables $\textbf{x}^{\mathcal{C}}$ and discrete variables $\textbf{x}^{\mathcal{D}}$. It represents the design choices, such as the task execution times, periods, and priority assignments. 
$\mathcal{F}(\textbf{x})$ maps the design variables $\textbf{x}$ into the design objective (a scalar). In addition, real-time system design is subject to the schedulability analysis constraint:
\begin{equation}
    \text{Sched}(\textbf{x})=\begin{cases}
        0, & \text{system is schedulable}\\
        1, & \text{otherwise}
    \end{cases}
    \label{general_sched_function}
\end{equation}
Here, we treat the schedulability analysis as a black-box that only returns binary results, i.e., schedulable or not.

\section{Methdology}
\begin{figure}[t]
\centering
\includegraphics[width=0.3\textwidth]{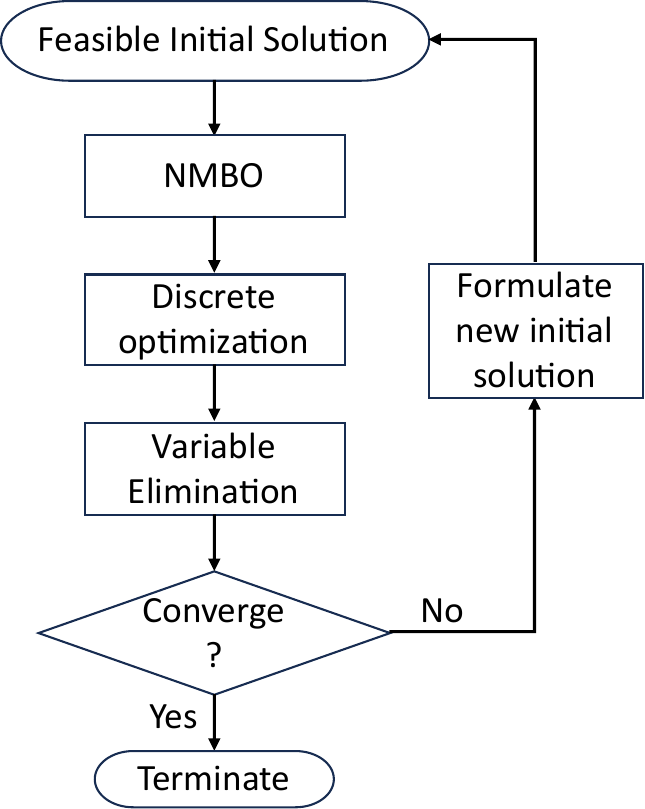}
\caption{NORTH+ is an extension of the NORTH optimization framework~\cite{Wang2023RTAS}. The NMBO and VE components are shown in Fig.~\ref{main_framework_fig_north} and explained more in Section~\ref{section_north_review}, the discrete optimization is introduced in Section~\ref{section_discrete_opt}. The continuous and discrete variables in NORTH+ are optimized separately: when optimizing continuous variables, the discrete variables are treated as constants, and vice versa. 
}
\label{fig_overview_co_design}
\end{figure}

The overview of NORTH+ is shown in Fig.~\ref{fig_overview_co_design}. 
 
\begin{figure}[ht]
\centering
\includegraphics[width=0.52\textwidth]{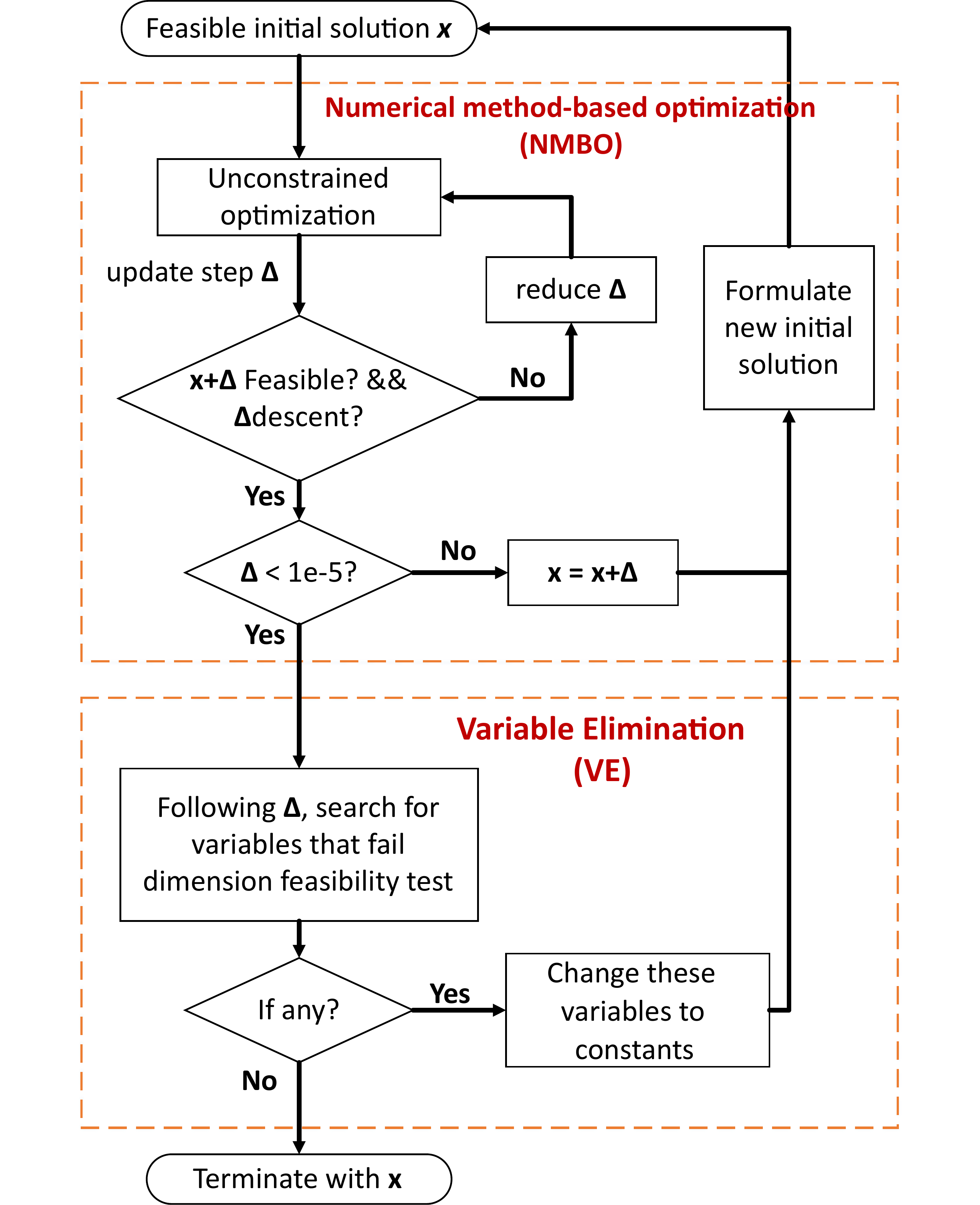}
\caption{Optimization framework NORTH from \cite{Wang2023RTAS} that optimizes only continuous variables.} 
\label{main_framework_fig_north}
\end{figure}

\subsection{Performing continuous optimization}
\label{section_north_review}
We briefly review \cite{Wang2023RTAS} in this section, which forms the foundation of this paper. The framework of \cite{Wang2023RTAS} is  shown in Fig.~\ref{main_framework_fig_north}, which has two major steps during iterations:
\begin{itemize}
    \item Numerical Method-Based Optimization (NMBO): an optimization method based on trust-region optimizers~\cite{Nocedal2006NumericalO2}. Given an initial solution $\textbf{x}$, the algorithm first performs unconstrained optimization to find a descent direction $\boldsymbol{\Delta}$ and then adjusts $\boldsymbol{\Delta}$ to keep $\textbf{x}+\boldsymbol{\Delta}$ feasible. It terminates when $\|\boldsymbol{\Delta}\|_2$ becomes smaller than a given threshold.
    \item Variable Elimination (VE): After NMBO terminates, we try to find a feasible descent direction (i.e., we can take a big step along this direction without violating the schedulability constraints) by searching within the sub-space of the $\boldsymbol{\Delta}$ above.
    When $\textbf{x}+\boldsymbol{\Delta}$ violates the schedulability constraints,  it is often the case that only a small subset of tasks miss their deadline. 
    Therefore, excluding these tasks from further optimization frees the numerical optimizers from schedulability constraints and improves performance in future iterations.
    \item Formulate new initial solution: Eliminated variables will be considered constant in subsequent iterations, and a new initial solution will be formulated based solely on the remaining variables.
\end{itemize}

\subsection{Performing discrete optimization}
\label{section_discrete_opt}

\begin{figure}[t]
\centering
\includegraphics[width=0.4\textwidth]{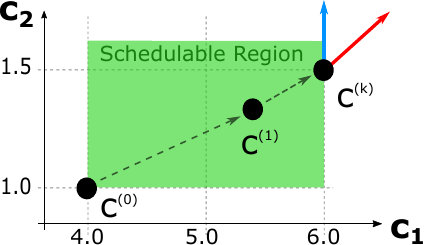}
\caption{ From \cite{Wang2023RTAS}, an example that visualizes how NORTH optimizes the continuous variables $\textbf{C}$ in the figure. Expanding the schedulable region at $\textbf{C}^{(k)}$ gives NORTH more chances of optimization.
}
\label{example_elimination}
\end{figure}

Although the primary goal of NORTH+ is optimizing the objective function~\eqref{general_F}, enhancing system schedulability can be regarded as a secondary goal. 
When NORTH optimizes the continuous variables, it tends to push them towards the boundaries of schedulable regions. An example is shown in Fig.~\ref{example_elimination}.
Thus, some variables are often close to violating the schedulability constraints when the numerical optimizer terminates. This situation prevents further optimization opportunities.  

Therefore, we can broaden the system's schedulability region to improve performance. 
For example, when the discrete variables are priority assignments, many classical algorithms can be utilized to assign priorities, such as Rate Monotonic (RM)~\cite{Pillai2001RealtimeDV, Bini2009MinimizingCE} or DkC~\cite{Davis2009PriorityAF}. 
These algorithms should be executed within each iteration, as illustrated in Fig.~\ref{fig_overview_co_design}, as optimal priority assignments may change as the continuous variables are optimized.

\begin{figure*}[hbt!]
    \centering
    \begin{subfigure}[b]{0.48\textwidth} 
        \centering
        \includegraphics[height=0.8\linewidth]{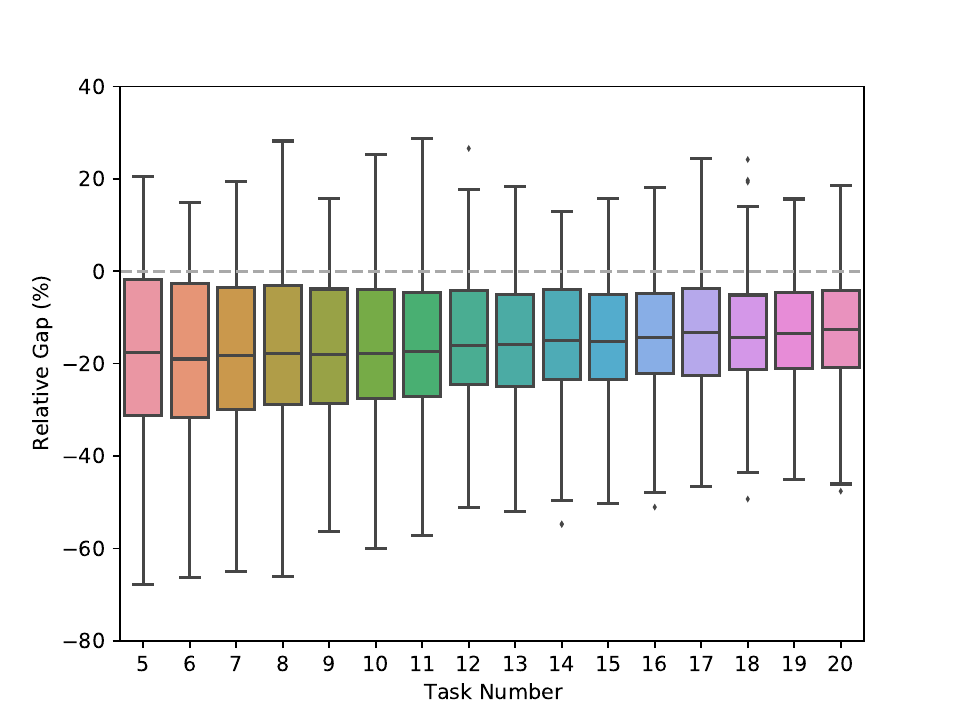}
        \caption{Relative performance gap}
        \label{exp_fig_rm}
    \end{subfigure}
    \begin{subfigure}[b]{0.48\textwidth}
        \centering
\includegraphics[height=0.8\linewidth]{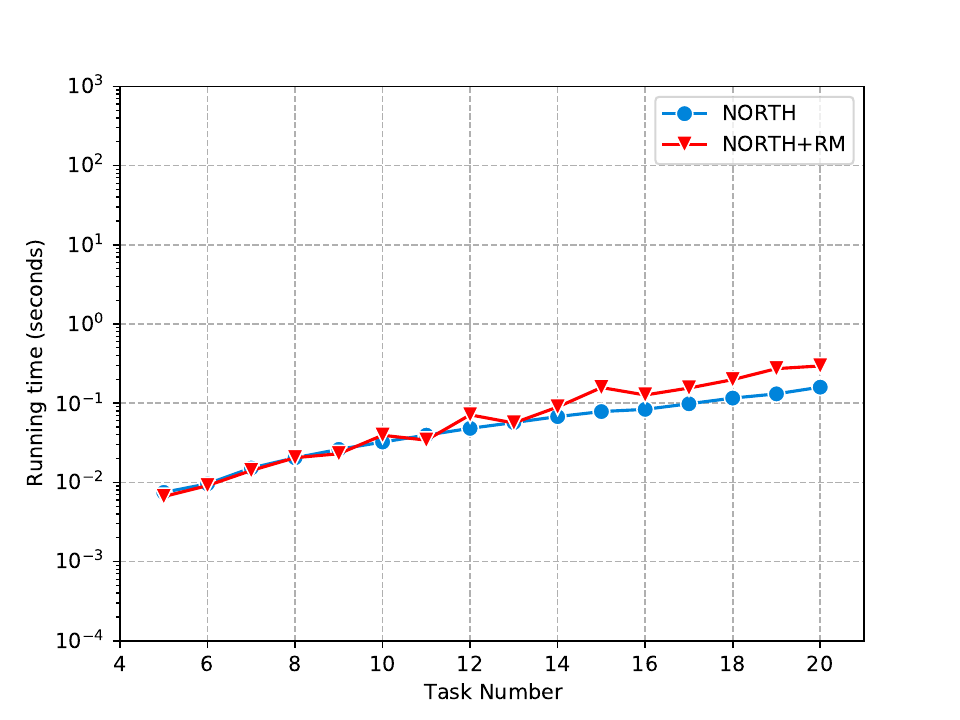}
    \caption{Run-time speed}
    \label{exp_fig_speed}
    \end{subfigure}
    \caption{Performance comparison between NORTH and NORTH+RM.}
    \label{control_all}
\end{figure*}

\section{Experiments}
In the experiments, we consider an example application where the objective function is derived from control performance optimization~\cite{Zhao2020AnOF}:
\begin{align}
    & \min_{\textbf{T},\textbf{P}} \sum_{i=0}^{N-1} (\alpha_i \textbf{T}_i + \beta_i r_i(\textbf{T},\textbf{P}))
    \label{ls_control}
\end{align}
where the variables $\textbf{T}_i$ are the task $\tau_i$'s periods; $N$ dentoes the number of tasks; $\alpha_i$ and $\beta_i$ are control weight parameters; $\textbf{P}$ are the priority assignments for all the tasks; $r_i(\textbf{T},\textbf{P})$ is the task $\tau_i$'s response time.

We evaluate the framework on simulated task sets. Ten thousand task sets were generated randomly following \cite{Zhao2020AnOF, Wang2023RTAS}. The task sets consisted of periodic tasks $\boldsymbol{\tau}$ with implicit deadlines scheduled by fixed-task priority schedulers on a uni-processor platform. 
The worst-case execution time $C_i$ of each task $\tau_i$ was generated randomly from [1, 100]; the task periods were upper-bounded by $T_i \leq 5 \sum_{i=1}^n C_i$. In the objective function \eqref{ls_control}, $\alpha_i$ was randomly generated in the range [1, 1000]; $\beta_i$ was generated in the range [1, 10000].

In the experiments, the black-box schedulability analysis constraint follows the classical response time analysis~\cite{Joseph1986FindingRT}:
\begin{equation}
    r_i = C_i +\sum_{j \in \text{hp}(i)} \ceil{\frac{r_i}{T_j}}{C_j}
    \label{rta_LL}
\end{equation}
where $\text{hp}(i)$ denotes the tasks with higher priority than the task $\tau_i$. Let $D_i$ denote the deadline of the task $\tau_i$, we have:
\begin{equation}
        \text{Sched}(\textbf{T}, \textbf{P})=\begin{cases}
        0, & \forall i, r_i(\textbf{T}, \textbf{P}) \leq D_i \\
        1, & \text{otherwise}
    \end{cases}
    \label{sched_model_ll}
\end{equation}

We compared the performance between NORTH~\cite{Wang2023RTAS} and NORTH+RM because they work with black-box schedulability constraints. 
In NORTH+RM, task priorities are adjusted with RM each time after NORTH optimizes the period variables, as illustrated in Fig.~\ref{fig_overview_co_design}.
Fig.~\ref{control_all} shows experiment results, where the relative gap is defined as follows:
\begin{equation}
    \frac{\mathcal{F}_{\text{NORTH+RM}}-\mathcal{F}_{\text{NORTH}}}{\mathcal{F}_{\text{NORTH}}} \times 100 \%
    \label{experiment_criteria}
\end{equation}

\section{Discussion, Open Challenges and Future Work}
Experiments show that adding an extra discrete optimization step to NORTH improves applicability and performance.
Although finding the optimal solutions for the discrete optimization step in NORTH+ could be difficult, simple heuristics may work reasonably well in practice. In simulated experiments, assigning priorities based on RM brought around 20$\%$ performance improvements. 

\subsection{Challenge: Optimality vs Applicability and Efficiency}
Fig.~\ref{exp_fig_rm} shows that NORTH+RM does not consistently outperform NORTH. NORTH+ is not designed to provide optimal solutions. Given a real-time system optimization problem with possibly black-box constraints and many variables, NORTH+ can provide a solution quickly with reasonably good quality, albeit without a guarantee of optimality.

Providing an optimality guarantee for problems with black-box constraints is challenging. These black-box constraints could be non-convex and non-continuous, and cause the optimal solutions, denoted as $\textbf{x}^*$, to potentially exist anywhere. 
In worst-case scenarios, many solution candidates we searched for may be infeasible or non-optimal until we find $\textbf{x}^*$, regardless of the optimization algorithms. In this case, enumerating all the possible solution candidates may be necessary to provide the optimality guarantee. This detrimentally impacts algorithm scalability.

If special properties of the black-box constraints are known, such as continuity or convexity, we can efficiently narrow down the search space. Although the problem is becoming less ``black-box'', domain-specific observation and techniques may be available to improve efficiency and potentially provide an optimality guarantee, such as sustainable schedulability analysis~\cite{Zhao2020AnOF}, convexity~\cite{Aydin2006SystemLevelEM}, dynamic programming~\cite{huang2023RTailor}. However, the availability of these domain-specific observations also implies a more limited range of applicability.

Another promising direction that aligns with the black-box assumption (i.e., without reducing applicability) is machine learning (ML) methods~\cite{Lee2021MLFR, Bo2021DevelopingRS, Wang2020RobotCU}. ML can construct mathematical models to approximate the black-box constraints and infer desirable properties. Once such a model is established, classical optimization algorithms can be employed to find good solutions. However, a significant challenge with these methods is ensuring feasibility, i.e., guaranteeing that the solutions found are always feasible.

\subsection{Future Work in Methdology Generalization}
\subsubsection{Multi-objective optimization}
Classical techniques from multi-objective optimization can be utilized with NORTH+ if there are multiple objective functions. These techniques include assigning weight parameters to combine multiple objectives into a scalar value or transforming specific objective functions into constraints.
For example, an objective function aiming to minimize response time could be transformed into a constraint that requires the response time to be smaller than a threshold. NORTH+'s general applicability works with both strategies mentioned above.

\subsubsection{Discrete optimization}
In this paper, we find that improving the system schedulability can indirectly improve the overall performance of NORTH. 
This observation encourages combining discrete optimization methods and schedulability analysis to achieve better performance. We leave this part as the future work.

\section{ACKNOWLEDGMENT}
This work is partially supported by NSF Grants No. 1812963 and 1932074.

\bibliographystyle{ieeetr}
\bibliography{scheduling}

\end{document}